\documentclass{aa}

\usepackage{graphicx}
\usepackage{color}
\usepackage{txfonts}
\newcommand{\bd}{BD+36$^\circ$3317 }
\newcommand{\bde}{BD+36$^\circ$3317}
\newcommand{\dlc}{$\delta$~Lyr cluster }
\newcommand{\dle}{$\delta$~Lyr cluster}
\newcommand{\phoebe}{\hbox{\tt PHOEBE }}
\newcommand{\phoebee}{\hbox{\tt PHOEBE}}

\newcommand{\korel}{\hbox{\tt KOREL }}
\newcommand{\korele}{\hbox{\tt KOREL}}

\newcommand{\spefo}{\hbox{\tt SPEFO }}
\newcommand{\spefoe}{\hbox{\tt SPEFO}}
\newcommand{\spel}{\hbox{\tt SPEL }}
\newcommand{\spele}{\hbox{\tt SPEL}}
\newcommand{\iraf}{\hbox{\tt IRAF }}

\newcommand{\ubv}{\hbox{$U\!B{}V$}}
\newcommand{\bv}{\hbox{$B\!-\!V$}}
\newcommand{\ub}{\hbox{$U\!-\!B$}}

\newcommand{\p}{$\pm$}

\newcommand{\m}{$^{\rm m}\!\!.$}

\newcommand{\kms}{km~s$^{-1}$ }
\newcommand{\ks}{km~s$^{-1}$}

\newcommand{\lgg}{{\rm log}~$g$ }

\newcommand{\rs}{R$_{\odot}$}

\newcommand{\ha}{H$\alpha$ }

\begin{document}

   \title{The orbital elements and physical properties of
the eclipsing binary \bde, a probable member of \dlc}

   \author{E. K\i ran
          \inst{1,2}
          \and
          P. Harmanec\inst{1}
          \and
    \"{O}. L. Değirmenci\inst{2}
  \and
  M.~Wolf\inst{1}
  \and
  J.Nemravov\'a\inst{1}
  \and
  M.~\v{S}lechta\inst{3}
  \and
  P.~Koubsk\'y\inst{3}
  }

   \institute{Astronomical Institute of the Charles University, Faculty of
   Mathematics and Physics, \hfill\break
   V~Hole\v{s}ovi\v{c}k\'ach~2, CZ-180 00 Praha~8 - Troja, Czech Republic
\and
   University of Ege, Department of Astronomy \& Space Sciences,
              35 100 Bornova - {\.I}zmir, Turkey
\and
   Astronomical Institute, Academy of Sciences of the Czech Republic,
   251 65 Ond\v{r}ejov, Czech Republic
        }

   \date{Received \today}


  \abstract
   {The fact that eclipsing binaries belong to a stellar group is useful, because the former can be used to estimate distance and additional properties of the latter, and vice versa.
}
 {Our goal is to analyse new spectroscopic observations of
BD$+36^\circ3317$ along with the photometric observations from the literature
and, for the first time,  to derive all basic physical properties of this
binary. We aim to find out whether the binary is indeed a member of
the $\delta$~Lyr open cluster.
}
   {The spectra were reduced using the IRAF program and the radial velocities were
measured with the program SPEFO. The line spectra of both components were disentangled with the program \korel and compared to a grid of synthetic spectra. The final combined
radial-velocity and photometric solution was obtained with
the program PHOEBE.
}
   {We obtained the following physical elements of \bde: $M_1$ = $2.24\pm0.07$ $M_{\sun}$, $M_2$ = $1.52\pm0.03$ $M_{\sun}$,
$R_1$ = $1.76\pm0.01$ $R_{\sun}$, $R_2$ = $1.46\pm0.01$ $R_{\sun}$,
$\log L_1$ = $1.52\pm0.08$ $L_{\sun}$, $\log L_2$ = $0.81\pm0.07$ $L_{\sun}$.
We derived the effective temperatures
$T_\mathrm{eff,1}$ = $10450\pm420$ K, $T_\mathrm{eff,2}$ = $7623\pm328$ K.
Both components are located close to ZAMS in the Hertzsprung-Russell (HR) diagram and their
masses and radii are consistent with the predictions of stellar evolutionary
models. Our results imply the average distance to the system
$\overline{d}$ = $330\pm29$ pc. We re-investigated the membership
of BD$+36^\circ3317$ in the $\delta$~Lyr cluster and confirmed it.
The  distance to BD$+36^\circ3317,  $ given above, therefore represents
an accurate estimate of the true distance for \dle.
}
{The reality of the $\delta$~Lyr cluster and the cluster membership of BD$+36^\circ3317$ have been reinforced.}


\titlerunning{Basic physical elements of \bde}
\authorrunning{K\i ran et al.}

   \maketitle
%

\section{Introduction}
Eclipsing binaries have played an important role in astrophysics
and in understanding  the nature and evolution of binary systems by
providing the most accurate values of stellar masses, radii, and luminosities.
Especially useful are eclipsing binaries, which are members of some kind of
cluster or association since they provide an excellent tool for accurately estimating
 the cluster distance and age, independently of photometric
calibrations.

The eclipsing binary \bd (GSC 2651 802, SAO 67556,
$\alpha_{2000} = 18^{\rm h}54^{\rm m}22^{\rm s}$,
$\delta_{2000} = 36^\circ51\farcm07\farcs445$, $V$ = 8\m77)  is located in the field of \dle. \citet{Step59}, who actually
discovered the $\delta$~Lyr (Stephenson~1) cluster, gives a visual magnitude
8\m8 and spectral type A0 for \bde. \citet{bron63} obtained photoelectric
\ubv\ and photographic observations of many stars in the vicinity of
$\delta$~Lyr and challenged the existence of the cluster. For \bd he
obtained $V=8$\m800, $(\bv)=+0$\m041, and $(\ub)=-0$\m036. However,
\citet{eggen68} made photoelectric (\ubv) photometry of 77 stars in the vicinity
of the \dlc and presented some convincing evidence that the cluster exists.
He derived the mean reddening $E(\bv)=0$\m05 and a distance modulus of 7\m5.
He argued that the \dlc has a similar colour magnitude and proper motion to
the Pleiades moving group. For \bd he obtained $V=8$\m80, $(\bv)=+0$\m02,
and (\ub)$=-0$\m08. \citet{eggen72} further developed the idea that several
clusters, including \dle, belong to the Pleiades moving group. For \bd he gave
$V_0=8$\m65, $(\bv)_0=-0$\m03, and $(\ub)_0=-0$\m115 and a spectral class
B9.5V. Later, \citet{eggen83} obtained $uvby$ photometry of stars from
the \dlc and mentioned that \bd is a spectroscopic binary with a radial
velocity (RV) range from $-90$ to $+17$~\ks. He gives $V=8$\m79,
$(b-y)=0$\m031, $m_1=0$\m150, and $c_1=0$\m885 for the system. These can
be compared to independent $uvby$ photometric results published by
\citet{ant84} as $V=8$\m90, $(b-y)=0$\m011, $m_1=0$\m160, and $c_1=0$\m904.
She obtained a large scatter in the distance moduli of individual cluster
members and again cast some doubt as to the existence of the cluster.
She confirmed, however, that the observed colours of \bd are indicative of
an A-type spectroscopic binary. Interestingly, neither  author noted that
the range of published values for the $V$ magnitude of \bd suggests its
light variability. However, in 2008,  \citet{vio2008} publish their
2007 $V$ band observations of the system and announce that \bd is
an eclipsing binary with a period of 4\fd30216. They also give the epoch of
the primary minimum as HJD~2454437.25921. \citet{ozdar2012} obtained a set
of complete \ubv\ light curves and improved the ephemeris to
\smallskip\noindent
\begin{equation}
T_{\rm min.I}={\rm HJD}\,2454437.2466(30)+4\fd302162(27)\times E\,.\label{efe}
\end{equation}
\smallskip\noindent
They derived a simultaneous solution of the light curves (LCs) and noticed
that the system has a total eclipse in the secondary minimum. As a result, they derived the magnitudes and colours of the components
separately. They give the intrinsic visual magnitudes and colours of
the components as $V_0=8$\m883, $(\ub)_0=-0$\m170, $(\bv)_0=-0$\m062
for the primary, and $V_0=10$\m277, (\ub)$_0=-0$\m104, (\bv)$_0=0$\m245
for the secondary. They estimate the interstellar reddening and total
visual extinction for the system as $E(\bv)=0$\m07, $A_0=0$\m22. From their
LC analysis, they estimate the absolute physical parameters of the components
and arrive at a distance of 353~pc for the binary. They do not give an error bar for the distance of the system.


\section{Observations and data reductions}
Spectroscopic observations of \bd were made with the single order
spectrograph attached to the 2~m reflector of the Ond\v{r}ejov Observatory,
Czech Republic. The spectra were recorded with a $CCD$ detector and cover the wavelength range of
$6260-6700$~\AA\ with a two-pixel spectral resolution of 11700.
The typical exposure times were 90 min, and the ratio of signal to noise (S/N) varies between 80
and 200. The system was observed over 20 nights from
March to July 2014. Each night, flat field and bias exposures were
obtained and the Thorium-Argon (ThAr) comparison spectra were obtained before and after each
stellar exposure. The initial reductions (bias subtraction, flat fielding,
cosmic ray removal, and wavelength calibration) were carried out with
the program \iraf by M\v{S}. Rectification of the spectra and the RV
measurements of the stellar and selected telluric lines were carried
out with the program \spefoe, which has written by Dr.~J.Horn \citep{sef0} and further
developed by \citet{sko96} and Mr. J.~Krpata.


\section{Towards basic physical properties of the binary}
\subsection{Direct RV measurements}
As already mentioned, the object had been classified as an A-type star. This is
corroborated by our red spectra, which contain the \ha line with a sharp core
and very broad wings, \ion{Si}{ii} doublet at 6347 and 6371~\AA,\ and
several weaker metallic lines, mainly of \ion{Fe}{i}, \ion{Fe}{ii},
\ion{Ca}{i}, \ion{Ni}{i}, and \ion{Mg}{ii}. For direct RV measurements,
we used the three strongest lines (\ha core and the \ion{Si}{ii} doublet),
where both binary components were easily resolved. The measurements
were carried out with the \spefo program, in which one can slide
an image of the flipped line profile with respect to the direct one
onto the computer screen until a perfect match of the desired parts of the profile
is achieved \citep[see, e.g.][for details]{zarfin30, sef0}. All spectra were
independently reduced and measured by EK and PH  and, after verifying
that both sets of these independent measurements agree well with each other,
mean values for each spectrum were adopted, as  recorded in
Table~\ref{jourv}.

\onltab{
\begin{table}
\centering
\caption[]{Heliocentric RVs of \bd measured in \spefo (mean of three
spectral lines).}
\label{jourv}
\begin{tabular}{crrrrr}
\hline\hline\noalign{\smallskip}
HJD$-$2400000&RV$_1$&RV$_2$\\
             &(\ks) &(\ks) \\
\noalign{\smallskip}\hline\noalign{\smallskip}
56744.5862 &  -91.3\p1.2&  91.7\p1.2 \\
56746.4816 &   63.5\p0.7&-143.6\p0.2 \\
56764.4510 &   11.3\p0.2& -63.8\p2.0 \\
56765.4209 &  -89.4\p0.3&  90.5\p1.3 \\
56778.4182 &  -93.4\p0.3&  94.3\p0.4 \\
56782.5511 &  -86.4\p1.3&  76.4\p3.3 \\
56799.5499 &  -67.4\p0.2&  56.9\p1.5 \\
56815.5215 &   59.5\p0.6&-135.4\p0.0 \\
56816.3854 &  -24.5\p1.3&   -- \\
56817.4476 & -100.2\p0.3& 100.7\p0.6 \\
56819.3930 &   61.3\p0.8&-138.0\p0.2 \\
56822.4202 &  -61.3\p0.3&  44.0\p0.2 \\
56826.4377 &  -88.0\p0.1&  78.7\p1.4 \\
56827.5433 &   31.1\p0.6&    -- \\
56852.4845 &  -63.2\p1.5&  51.9\p1.0 \\
56852.5226 & -61.6\p0.2 &  47.9\p4.6 \\
56861.5258 & -14.6\p0.5 &    -- \\
56862.4931 &  62.4\p0.9 &-139.8\p0.0 \\
56865.4533 &  58.2\p0.3 &  41.0\p0.6 \\
56866.5585 &  52.1\p0.9 &  -115.5\p0.9 \\
\noalign{\smallskip}\hline\noalign{\smallskip}
\end{tabular}
\end{table}
}


\subsection{A trial RV solution with the program \spele}
To have some guidance for a more sophisticated analysis, we first derived
the orbital solution with the program \spele, written by the late Dr. J.~Horn.\footnote{The program has never been published but was carefully tested
and used in several publications. http://astro.troja.mff.cuni.cz/ftp/hec/SPEL90/spel.pdf} The program requires input values for
the orbital elements (orbital period and epoch of maximum RV,
semi-amplitudes of the radial velocity curves of the components, and
the systemic velocity of the binary in this case). We kept the orbital period
from ephemeris~(\ref{efe}) fixed and adopted a circular orbit. There seems
to be some weak evidence of a very small eccentricity of the orbit
but only continuing observations of the times of minima could (dis)prove
it. In any case, the use of the circular orbit has a negligible effect
on the elements that define the binary masses.
The results can be found in Table~\ref{spel90}.

\begin{table}
\centering
\caption[]{Orbital parameters and their uncertainties obtained with the program \spele.}
\label{spel90}
\begin{tabular}{rccc}
\hline\hline\noalign{\smallskip}
Element&Value\\
\noalign{\smallskip}\hline\noalign{\smallskip}
$P$  (d) &  4.30216 fixed\\
$T_{\rm min.I}$  (HJD) & $2454437.2359\pm0.0046$ \\
$K_1$  (\ks) & $82.6\pm0.6$\\
$K_2$  (\ks) & $123.1\pm0.7$\\
$q=K_1/K_2$  & $0.67\pm0.01$\\
$V_0$  (\ks) & $-19.0\pm0.4$ \\
$a \sin i$  (\rs) & 17.5 \\
\noalign{\smallskip}\hline\noalign{\smallskip}
\end{tabular}
\end{table}

\subsection{Improved linear ephemeris}
Having now two sets of photometric observations and
new RVs, which span a substantially longer time interval than before, we
decided to derive a new, more accurate linear ephemeris. To this end,
we first used the program \phoebe 1.0 \citep{prsa2005, prsa2006}, which
is an extension of the widely used WD program \citep{wd71}.
As mentioned above, \citet{vio2008} obtained the first $V$-band light curve
of the \bd and derived a linear ephemeris. Later, \citet{ozdar2012}
published a new ephemeris based on their \ubv\ light curves.
We combined all four LCs with our \spefo RVs to obtain the following
linear ephemeris:
\smallskip\noindent
\begin{equation}
T_{\rm min.I}={\rm HJD}\, 2454437.2480(13)+4\fd302152(1)\times E\,.\label{efe2}
\end{equation}
\smallskip\noindent


\subsection{Spectra disentangling and another trial solution}
To check  the results from the direct RV measurements in \spefoe, and
to obtain line spectra of individual binary components that were suitable for further
analyses, we decided to disentangle the spectra. For this
we used the program \korele, written and further developed by
\citet{had95, had97, had2004, had2009}, and \citet{sko2010},
with the latest version available through the VO-KOREL web service.
Before preparing the input data for \korele, we estimated the S/N
of individual spectra in the line-free region 6625 -- 6645~\AA\ as
the ratio of the mean signal and its rms error using the formula
\begin{equation}
S/N=\left(\frac{\sum S}{m}/
    \sqrt{\frac{\left(\sum S^2-(\sum S)^2/m\right)}{m-1}}\right),\label{sn}
\end{equation}
\noindent where $m$ is the number of pixels used in the wavelength interval. We then weighted each spectrum by the weight proportional to
(S/N)$^2$ and normalized to the mean S/N of all spectra that were used.
The rebinning of the electronic spectra and preparation of input data
for \korel was carried out with the program {\tt HEC35D} written by PH.\footnote {The program and a manual to it can be downloaded from \sl http://astro.troja.mff.cuni.cz/ftp/hec/HEC35\,.}

\begin{figure}
   \centering
   \includegraphics[scale=0.45,angle=270]{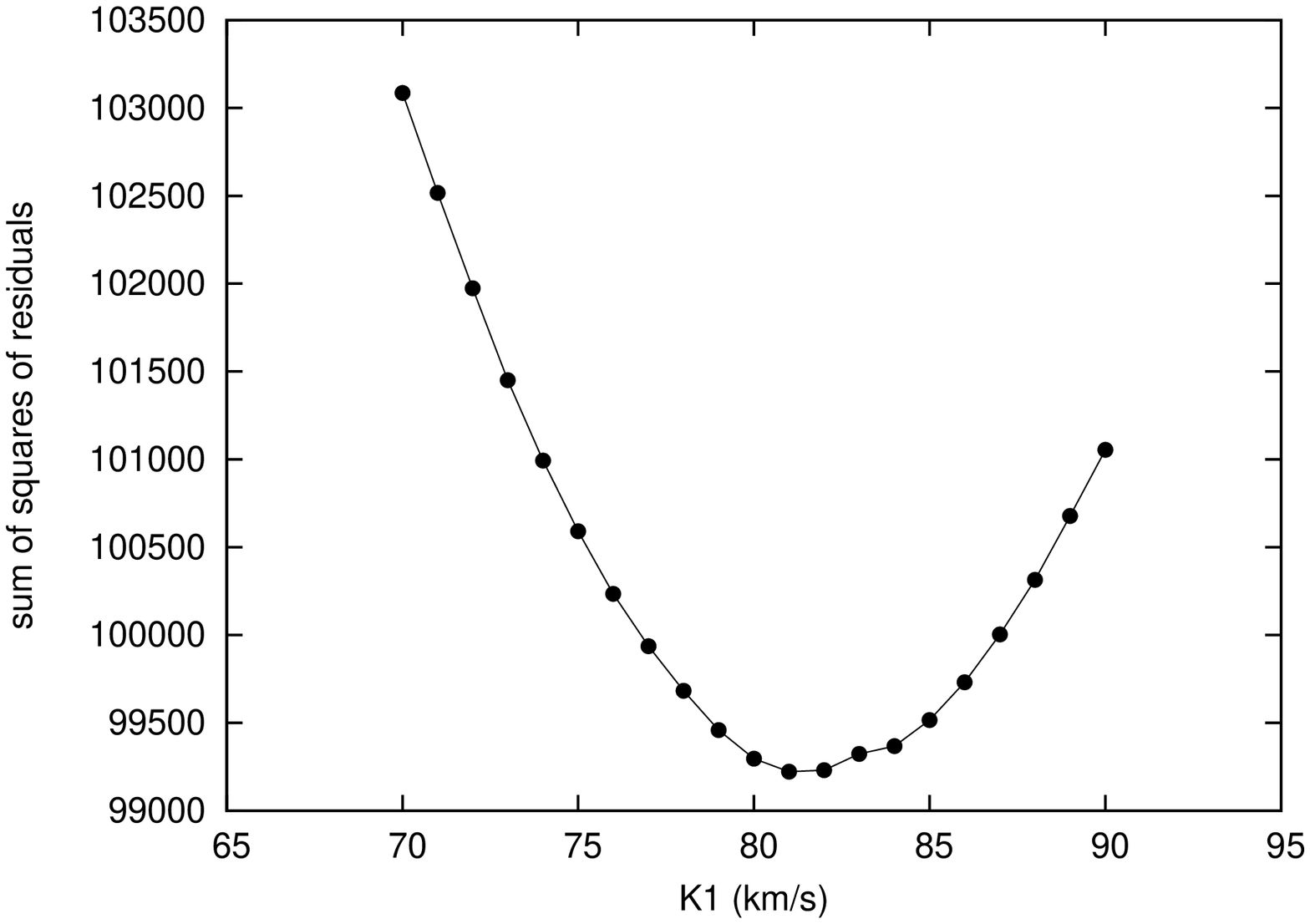}
   \includegraphics[scale=0.45,angle=270]{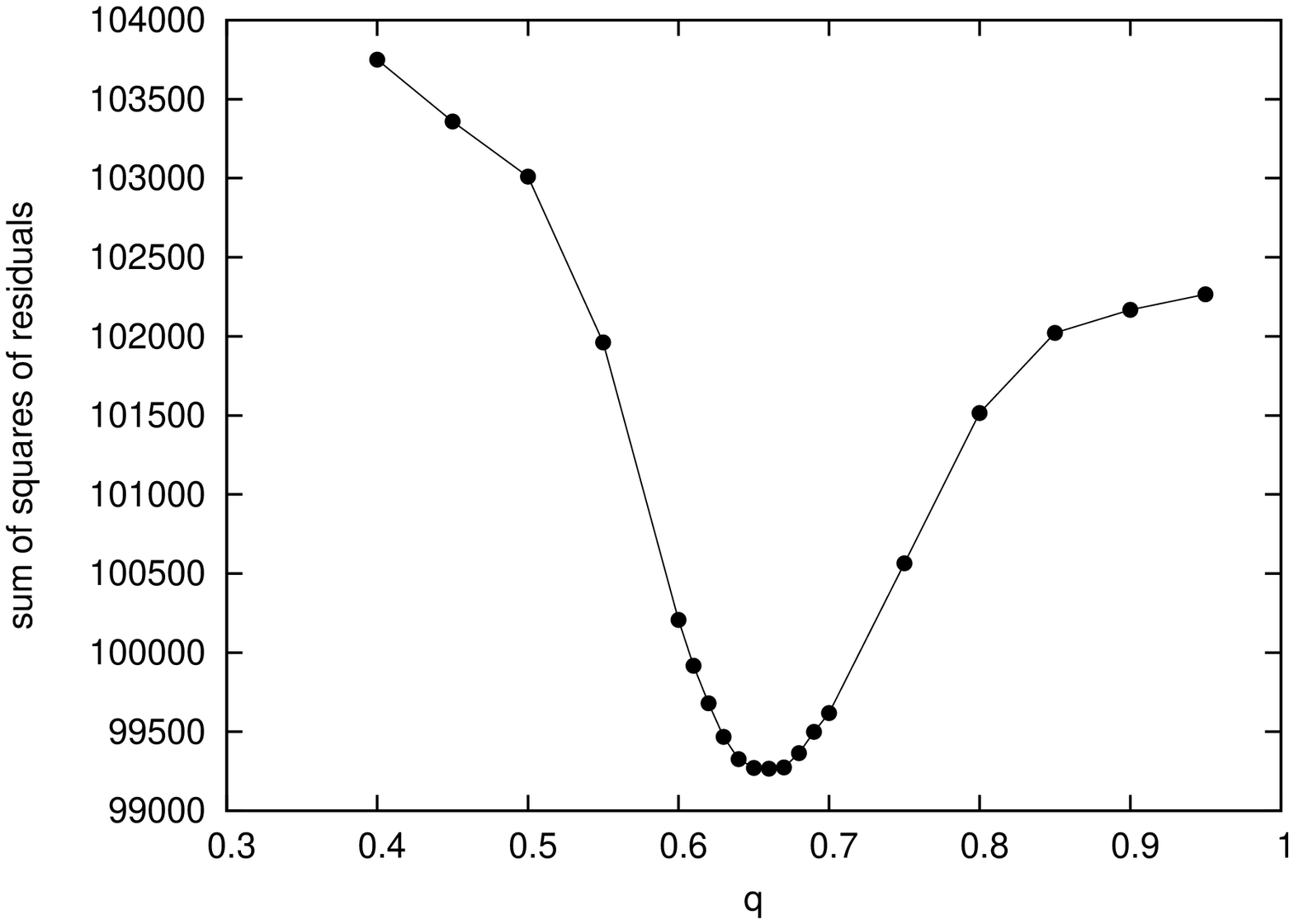}
      \caption{Sum of squares of residuals for trial \korel solutions
with fixed elements as a function of (a) the semi-amplitude $K_1$ (upper panel),
and (b) the mass ratio $K_1/K_2$ (lower panel).}
         \label{mapping}
\end{figure}

Using the ephemeris (\ref{efe2}) and the orbital parameters derived
with \spele, we run a number of tests. First,

we kept all elements fixed and ran the program for several different
values of the semi-amplitude $K_1$ to find out which one gives the lowest
sum of squares of the residuals. This indicated $K_1\sim81$~\kms
as the optimal value (see the upper panel of Fig.~\ref{mapping}). Keeping this
value fixed, we carried out a similar mapping to obtain the best guess for
the mass ratio $q = K_1 / K_2$. The result is shown in the bottom panel of
Fig.~\ref{mapping} ($q\sim0.66$).

\begin{figure*}
   \centering
   \begin{tabular}{cc}
   {\includegraphics[scale=0.45,angle=270]{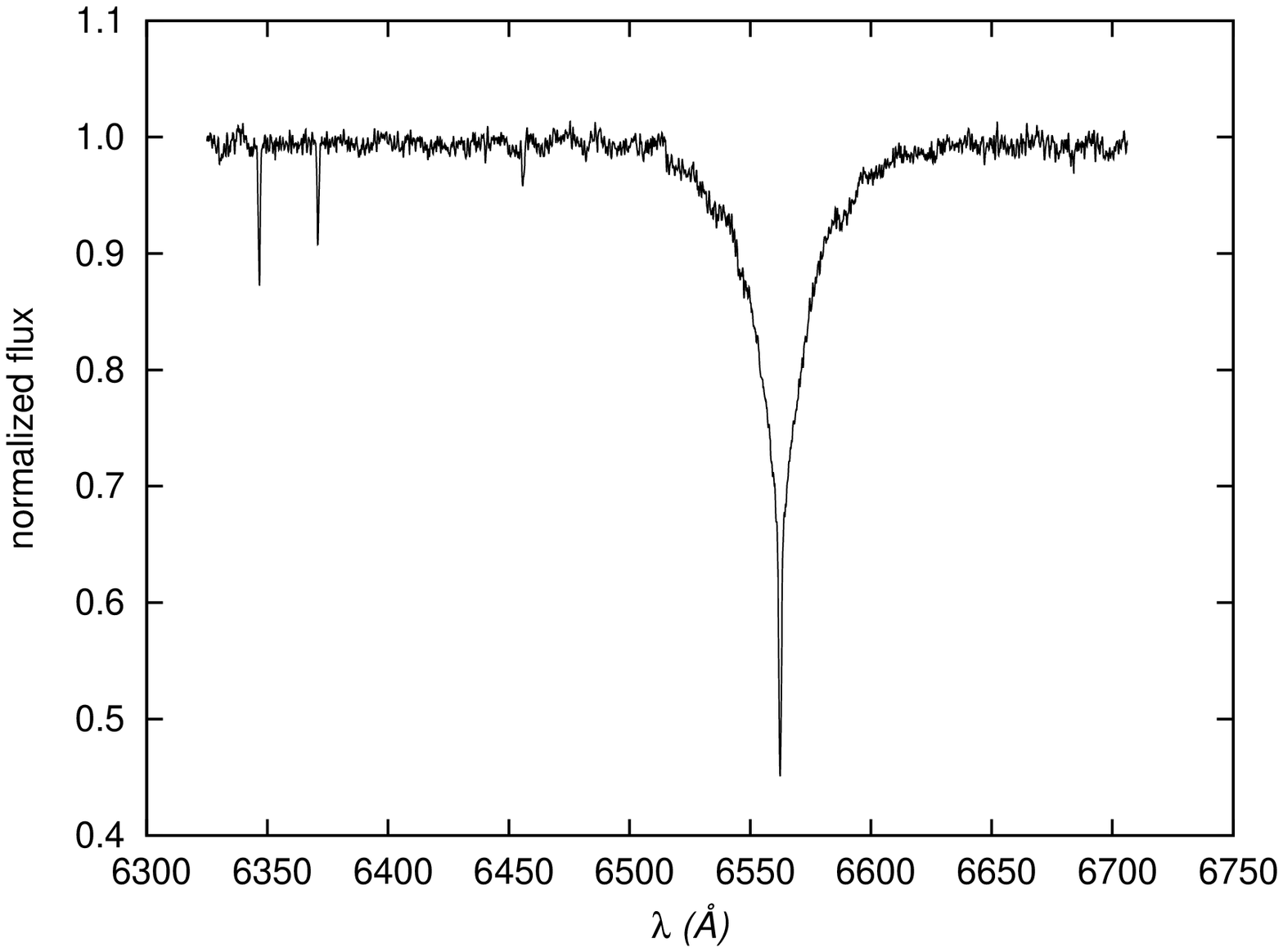}}&{\includegraphics[scale=0.45,angle=270]{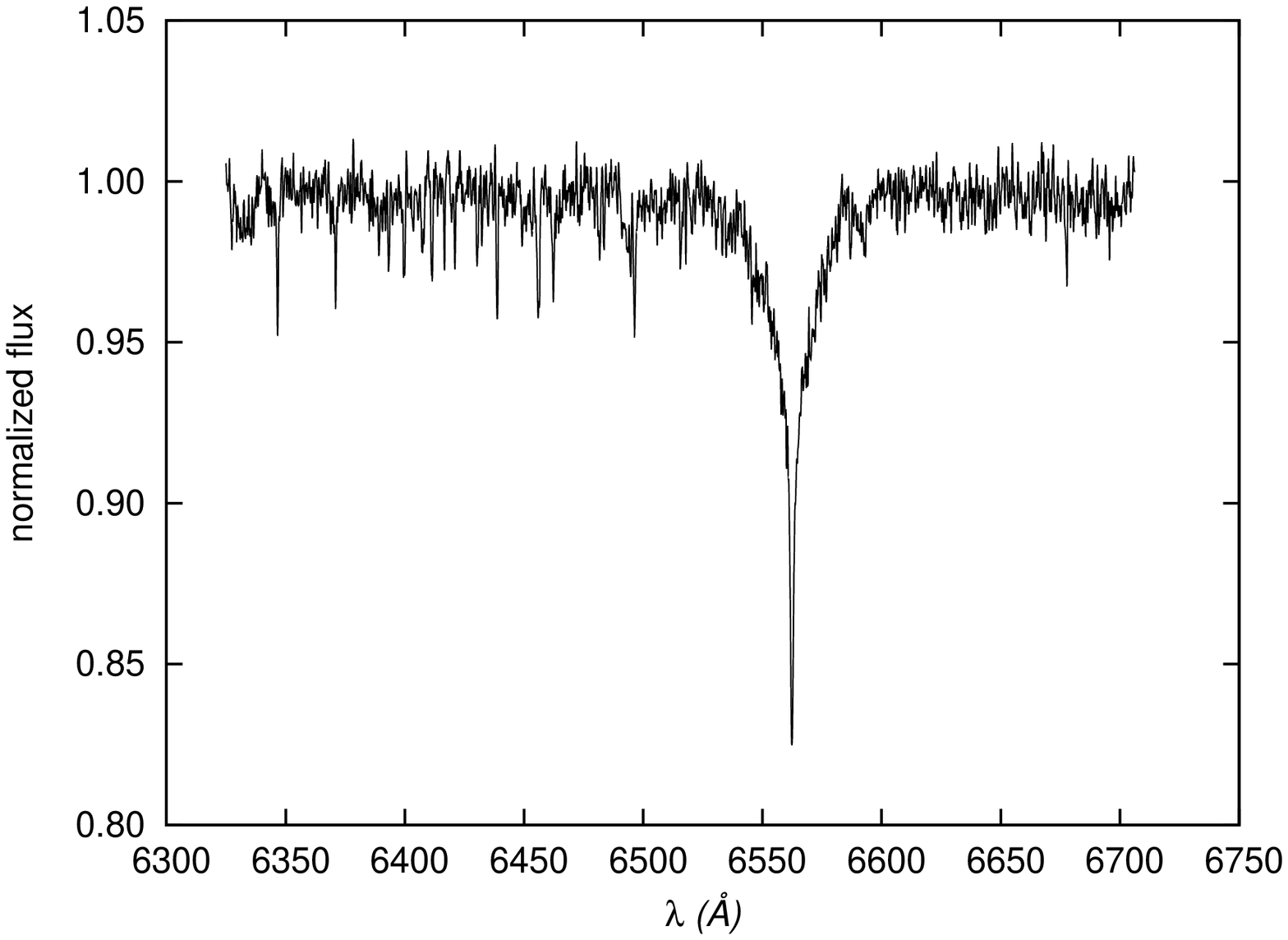}}\\
    \end{tabular}
    \caption{Disentangled spectra of primary (left panel) and secondary (right panel) components obtained with \korele.}
         \label{disenspec}
\end{figure*}

Starting with these values, we then carried out a series of \korel solutions,
but now allowing the free convergence of the epoch $T_{\rm min.I}$, semi-amplitude
$K_1$, and the mass ratio $q=K_1/K_2$ and variously kicking away the initial values
from the adopted values.
The solution that gives the lowest sum of squares of residuals are shown in
Table ~\ref{korelRV}, and the corresponding disentangled spectra of
the components are shown in Fig.~\ref{disenspec}.
We note that the \spel solution based on RVs measured in \spefo
does not differ significantly from the optimal \korel solution.

\begin{table}
\centering
\caption[]{Best trial orbital solution with \korel (see the text for the estimation procedure of the error bars).}
\label{korelRV}
\begin{tabular}{rccrrrrr}
\hline\hline\noalign{\smallskip}
Element&Value\\
\noalign{\smallskip}\hline\noalign{\smallskip}
$P$  (d) &  4.302152 fixed\\
$T_{\rm min.I}$  (HJD) &  2454437.26469  \\
$K_1$  (\ks)&  $80.8\pm1.1$ \\
$K_2$  (\ks)&  $124.1\pm1.3$ \\
$q$ &  $0.650\pm0.015$ \\
$a sini$  (\rs)&  17.4 \\
\noalign{\smallskip}\hline\noalign{\smallskip}
\end{tabular}
\end{table}

  The  \korel program does not provide error estimates of individual
parameters. However, certain error estimates can be obtained
as follows.  First, since there is a good agreement between the results
from RVs directly measured with \spefo and those from \korele, one can
adopt the $K$ and $q$ errors from the \spel solution in Table 2 as being
quite representative of real uncertainties. Secondly, we can also estimate
uncertainties of ${K_1}$ and ${K_2}$ from the RV measurements
given in Table ~\ref{jourv}. According to the values in
Table ~\ref{jourv}, our RV measurements have mean errors of 0.63 and 1.1 \kms
for the primary and secondary components, respectively. These values are
also in  good agreement with the $K$ errors in Table~\ref{spel90}.
Thirdly, a close inspection of the rather flat minima in Fig.~\ref{mapping}
suggests errors of $\sigma_{K_1}$=2 \kms and $\sigma_{q}$= 0.02 for $K_1$
and $q$, respectively.
Therefore, finally, we  adopt the average errors of $\sigma_{K_1}$=1.1 \ks,
$\sigma_{K_2}$=1.3 \ks, and $\sigma_{q}$= 0.015 for $K_1$, $K_2$ and $q$,
respectively.


\subsection{A comparison of disentangled and observed spectra with
synthetic ones}
To obtain the estimates of the effective temperatures, gravity accelerations,
and projected rotational velocities from spectroscopy, we used two independent
procedures.

First we used a~program which compares synthetic spectra with disentangled or observed
spectra of multiple systems to estimate radiative properties of its
components  \citep[see][for the details]{nasseri2014}.
The program uses several pre-calculated grids of synthetic spectra.
The primary falls within parameters covered with
grid POLLUX~\citep{palacios2010} and the secondary falls
within grid AMBRE~\citep{delaverny2012}. The wavelength band
$\Delta\lambda\in\lbrace{6330-6695\rbrace}$\,\r{A} was fitted.
The optimized parameters were the effective temperature,
the~projected rotational velocity, the systemic radial velocity,
and the relative luminosity\footnote{The total luminosity was
constrained as follows: $L_{1}+L_{2}=1.0$.} of both components.
The result is presented in Column~2 of Table~\ref{specres}.

Uncertainties of fitted parameters were estimated with a
Monte Carlo simulation, which was carried out as follows:
1) The continuum $\sigma_{\rm c}$ noise was estimated
for each disentangled spectrum, using the same procedure as the previous section.
2) An artificial Gaussian  noise with $\sigma=\sigma_{\rm c}$
was added to the disentangled profile.
3) The adjusted disentangled spectrum was fitted.
This procedure was repeated 500 times and
the errors were estimated from the distribution of
results from all runs.
These uncertainties do not reflect the need for
the~re-normalization of disentangled profiles.
This step especially affects  the width of H\,$\alpha$
line and, consequently, the obtained parameters, particularly
the gravitational acceleration and the projected velocity.
The gravitational acceleration was estimated from
the light curve solution and fixed during the fitting of disentangled
spectra, but the projected rotational velocity was fitted.

To test the reliability of the estimated projected rotational
velocity, another fit that followed the same procedure was
computed, but the fitted wavelength range was only
$\Delta\lambda\in\lbrace{6342-6350; 6368-6374\rbrace}$\,\r{A}. This region contains a pair of silicon lines
\ion{Si}{II}{\,6347}\,\r{A} and \ion{Si}{II}{\,6371}\,\r{A}, which
should be unaffected by the re-normalization uncertainty.
The optimal rotational velocities are $v_1\sin i = 32.55 \pm 0.59$\,\ks, and
$v_2\sin i = 22.1 \pm 1.2$\,\ks. This shows that
the true uncertainty of the rotational velocity
of secondary is $\sim10$\,\ks.

As an independent approach to determine the effective temperatures of the components we used the program {\tt COMPO2}
written by \citet{Fras2006}, which combines two reference spectra for both components for given effective temperatures, radial velocities (or systematic velocity), projected rotational velocities, and gravitational accelerations and compare the combined spectrum to the observed spectrum of a binary system. To achieve this, we used the reference spectra taken from \citet{val2004} and tried to reproduce our observed spectrum taken at maximum orbital elongations (0.75 phase). {\tt COMPO2} tries to minimize the residuals between observed and composite spectra to find optimal parameter values for effective temperature and fractional flux contributions. The results of this procedure are listed in Column 3 of Table~\ref{specres}. As seen from the table, both methods estimate almost the same values for the primary's effective temperature. However the relative contributions of the components to the total luminosity in the spectral regions under consideration  do differ somewhat from each other.

 \begin{table}
 \centering
 \caption[]{Properties of the binary components of \bd estimated
from the comparison of the disentangled (Column~2) and observed (Column~3)
spectra with synthetic ones (see the text for details).}
 \label{specres}
 \begin{tabular}{c|ccc}
 \hline\hline\noalign{\smallskip}
 Elements&Disentangled & Observed\\
 \noalign{\smallskip}\hline\noalign{\smallskip}
 $T_\mathrm{eff,1}$  (K) & $10500\pm340$ & $10400\pm420$ \\
 $T_\mathrm{eff,2}$  (K) & $7180\pm850$ & $7600\pm880$ \\
 $\lgg_{1}$  [cgs] & $4.29^{1}$ & $4.31\pm0.07$ \\
 $\lgg_{2}$ [cgs] & $4.29^{1}$ & $4.29\pm0.04$ \\
 $L_1/L_\mathrm{tot}$ & $0.74\pm0.01^{2}$ & $0.80\pm0.02^{2}$ \\
 $L_2/L_\mathrm{tot}$ & $0.26\pm0.01^{2}$ & $0.20\pm0.02^{2}$ \\
 $v_1\sin i$ (\ks) & $29.1\pm1.1$  & -- \\
 $v_2\sin i$ (\ks) & $30.7\pm4$  & -- \\
 $\gamma_1$ (\ks) & $-18.1\pm0.2$ & -- \\
 $\gamma_2$ (\ks) & $-17.5\pm0.6$ & -- \\
 \noalign{\smallskip}\hline\noalign{\smallskip}
 \end{tabular}
 \tablefoot{$^1$The gravitational acceleration
 was fixed a value obtained from the light curve solution. $^2$These values represent the luminosity ratio for the studied spectral region.}
 \end{table}

\subsection{Simultaneous solution of light and radial velocity curves with \phoebee}\label{pho}
The final solution to obtain the binary masses, radii, and luminosities
was carried out with the program \phoebee. The RVs presented in Table~\ref{jourv} and all four LCs obtained from the literature (Johnson $V$ observations were taken from \citet{vio2008}, and the Johnson $U$, $B$, and $V$ observations from \citet[][and priv.com.]{ozdar2012}) are solved simultaneously.
For the temperatures of the components, the bolometric albedos $\it A_{1,2}$ and gravitational darkening
coefficients $\it g_{1,2}$ were taken from \citet{clar2001} and \citet{clar98},
respectively. The limb-darkening
coefficients $\it x_{1,2}$ were computed automatically by \phoebe from internal tables of the programme which were generated according to the models given by \citet{cas2004}. Convergence was allowed for
the epoch of primary minimum
$T_{\rm min.I}$, the orbital period $P$, the semi-major axis of the relative orbit {\it a},
systemic velocity {\it $V_\gamma$}, orbital inclination {\it i},
dimensionless surface potentials of the components {\it $\Omega_1$} and
{\it $\Omega_2$}, the effective temperature of the secondary component
$T_\mathrm{eff,2}$, mass ratio {\it q,} and relative monochromatic luminosities
of the primary component $L_1$ in individual photometric pass-bands.

\begin{figure}
   \centering
   \hspace*{-0.5cm}
    \includegraphics[scale=0.52,angle=270]{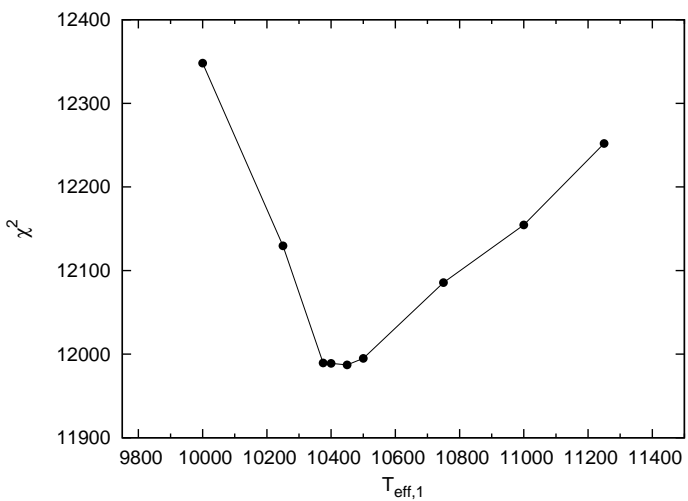}
      \caption{Search for the best value of the effective temperature ($T_\mathrm{eff,1}$) for primary component with \phoebe solutions.}
         \label{T1sear}
\end{figure}

To refine the temperatures of the primary component, which we obtained in previous sections, we applied an $T_\mathrm{eff,1}$ search procedure.
We subsequently solved the RV and LC curves simultaneously for a number
of fixed values of $T_\mathrm{eff,1}$ between 10000 K - 11250 K.
The run of the weighted sum of the square of residuals ($\chi^2$=$[\Sigma$W$(O-C)^2]$) thus obtained as a~function of the effective
temperature of the primary is shown in Fig ~\ref{T1sear}.
As can be seen from the figure, $\chi^2$ has its lowest value around $T_\mathrm{eff,1}$=10450 K which agrees with $T_\mathrm{eff,1}$=$10750\pm450$ K given by \citet{ozdar2012} within the uncertainty bars, and so for the final solution we adopted $T_\mathrm{eff,1}$=10450 K.

\begin{figure}
    \begin{center}

    \hspace*{-1.8cm}{\includegraphics[scale=0.57,angle=270]{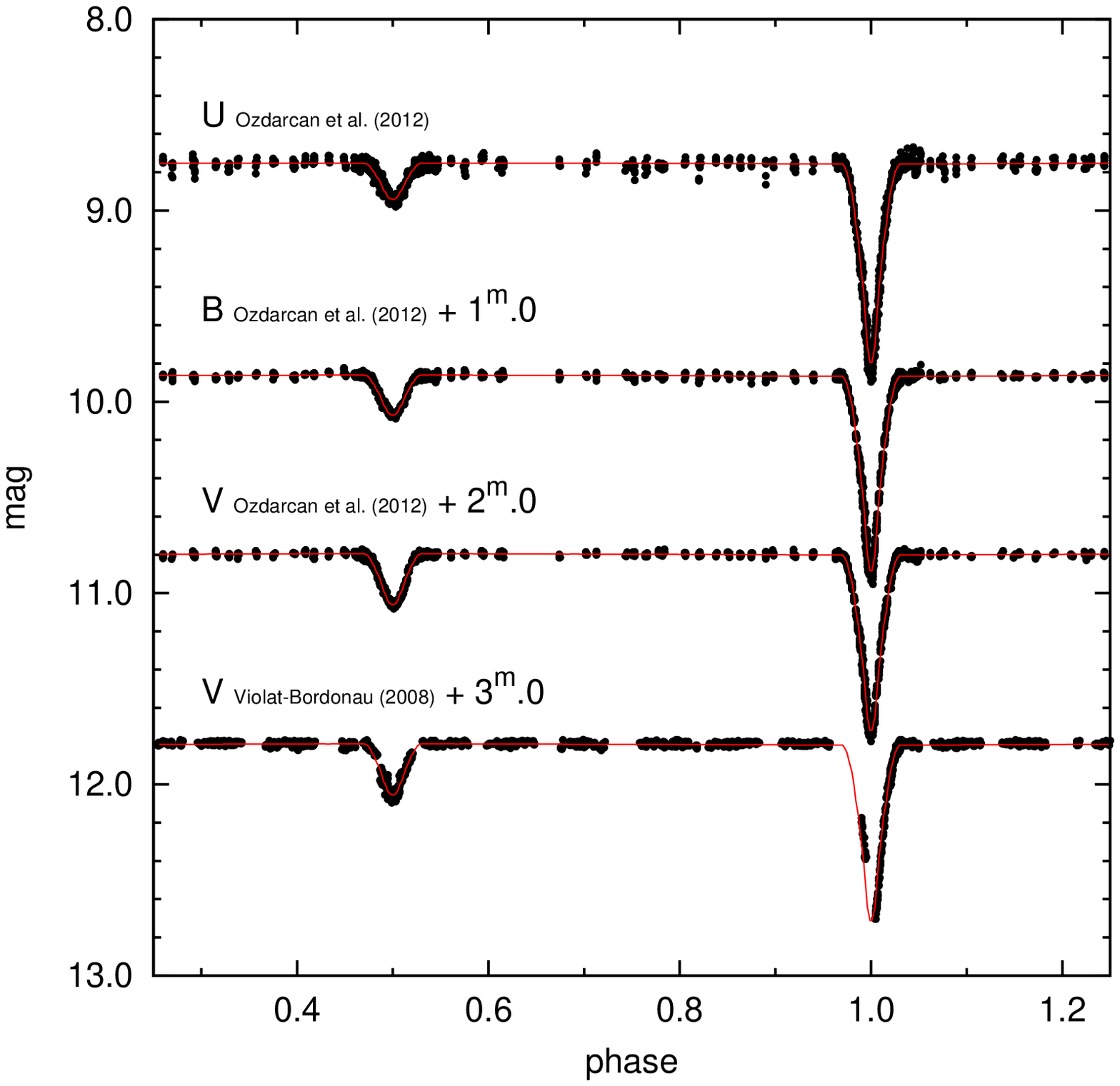}}\\
    \hspace*{-0.2cm}{\includegraphics[scale=0.51,angle=270]{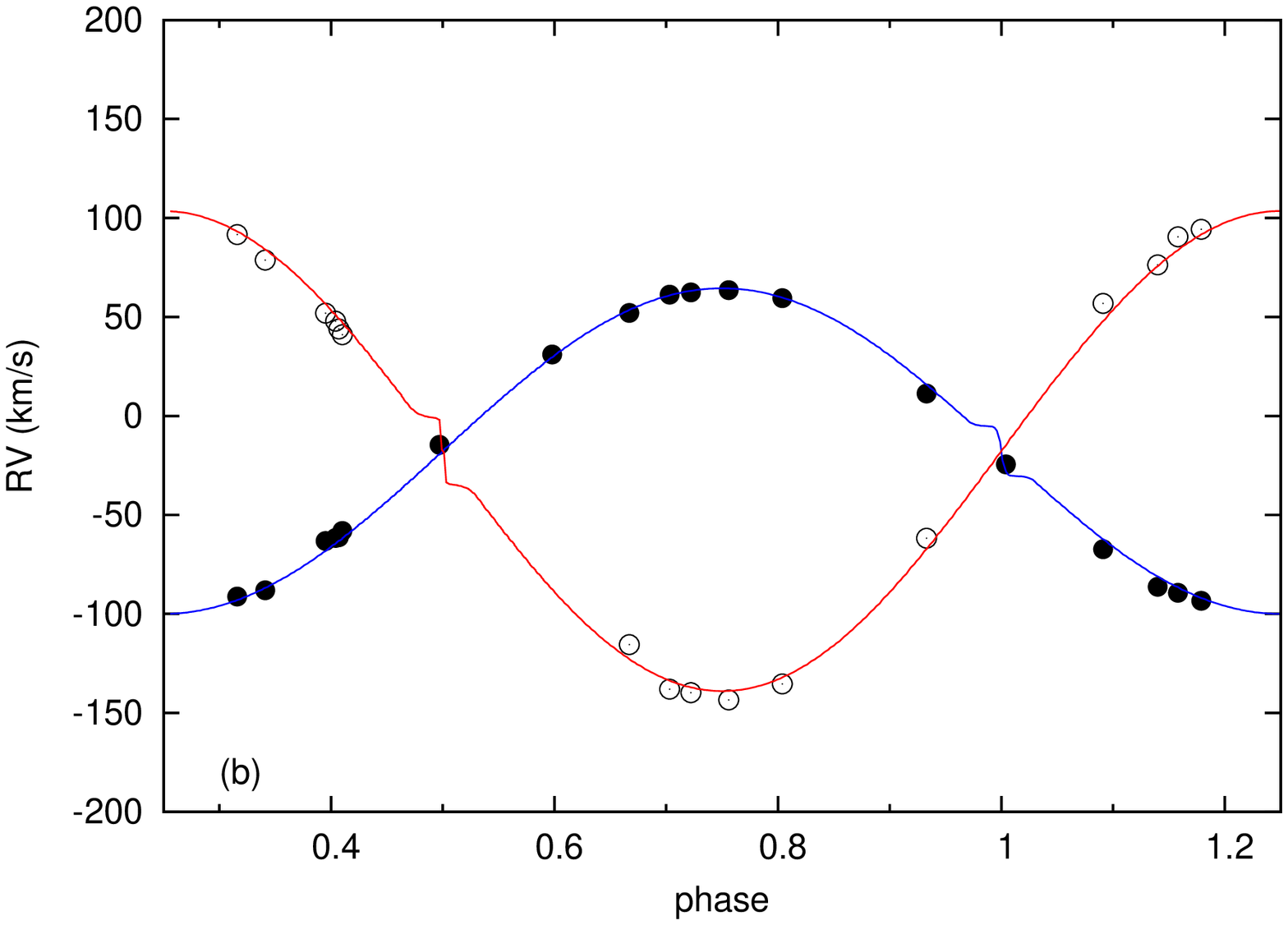}}

   \caption{(a) Observed (dots) and theoretical (solid lines) light curves of \bde. The theoretical
curves are calculated using the parameters given in Table~\ref{final}. (b) A phase plot of RVs of \bde. Black dots and open
circles denote the RVs of the primary and secondary, respectively.
Solid lines show the theoretical RV curves derived with the \phoebe program
(see Section~\ref{pho}). We note the presence of a small Rossiter
(rotational) effect.}
   \label{LCRV}
    \end{center}
\end{figure}

The final results are shown in Table~\ref{final}. The values of the parameters $P$, $i$, $a$, and $q$ given in Table~\ref{final} correspond to $K_1 = 82.23\pm0.82$ \ks and $K_2 = 121.29\pm0.85$ \ks, which match the results of \spel and \korel in previous subsections.
The model light and RV curves are compared with the observations in
Fig.~\ref{LCRV}. They show a small
Rossiter (rotational) effect \citep{rossit} in the RV curves
near the phases of binary eclipses.

\begin{table*}
    \centering
\begin{center}
\vspace{0.1cm} \caption[ ]{Final solution for the physical properties
of \bde. The uncertainties are derived locally by \phoebe from the covariance
matrix.}
\label{final}
\begin{tabular}{r|cccccc}
\hline\hline\noalign{\smallskip}
Element & Primary &System & Secondary \\
\noalign{\smallskip}\hline\noalign{\smallskip}
$T_{\rm min.I}$ (HJD) && $2456803.4598\pm0.0001$&\\
$P$ (d) & & $4.302152\pm0.000001$    &\\
$e$ & & 0.0 (fixed)    &\\
$a$ $(R_{\sun})$ && $17.3\pm0.1$ &\\
$V_\gamma$ (\ks) && $-17.8\pm0.5$ &\\
$x_{bol}$ & -0.02 && 0.14\\
$A$ & 1.00 && 0.92\\
$g$ & 1.00 && 0.90\\
$i$ $({}^\circ)$ && $89.27\pm0.02$&\\
$T_\mathrm{eff}$ (K) & 10450 (fixed)&& $7623\pm8.1$\\
$\Omega$  & $10.49\pm0.02$ && $9.19\pm0.02$\\
$q$ &   & $0.678\pm0.002$ &\\
$\lgg$ [cgs] & 4.29 && 4.29\\
$(l/l_{tot})^{1}$  $U$ band & $0.84\pm0.01$ & & 0.16\\
$(l/l_{tot})^{1}$  $B$ band & $0.83\pm0.01$ & & 0.17\\
$(l/l_{tot})^{1}$  $V$ band & $0.79\pm0.01$ & & 0.21\\
$(l/l_{tot})^{2}$  $V$ band & $0.79\pm0.01$ & & 0.21\\
$r_{pole}$     & $0.1019\pm0.0002$ && $0.0844\pm0.0002$ \\
$r_{point}$   & $0.1021\pm0.0002$ && $0.0846\pm0.0002$ \\
$r_{side}$     & $0.1020\pm0.0002$ && $0.0845\pm0.0002$ \\
$r_{back}$  & $0.1021\pm0.0002$ && $0.0846\pm0.0002$ \\
$(N)^{3}$   && 6847&\\
$\chi^2$   && 11987&\\
\noalign{\smallskip}\hline\noalign{\smallskip}
\end{tabular}
\tablefoot{$^1$ \citet{ozdar2012}, $^2$ \citet{vio2008}, $^3$ the total number of the observed (LC and RV) points.}
\end{center}
\end{table*}

\subsection{Physical properties of the system}
Using the results of the final solution from Table~\ref{final},
we calculated the basic physical properties of the components. These are summarised in Table~\ref{pyhs}. The masses of
the primary and secondary components are consistent with the spectral
types A3 V and F2 V, respectively. Using the observed {\it V}-band magnitude
outside the eclipses and (\ub) and (\bv) colours given by \citet{ozdar2012},
the relative luminosities from the \phoebe solution, and the bolometric
corrections  from \citet{gra2005}, we calculated the de-reddened magnitudes,
colours, distance moduli and the distances of the components. These are also
given in Table~\ref{pyhs}. Considering that the primary is much brighter
than the secondary, we weighted the individually derived distances to the components with their luminosities and estimated the mean distance to the system as $\overline{d}=330\pm29$~pc.

\begin{figure}
\hspace*{-2cm}
\includegraphics[angle=270,scale=0.55]{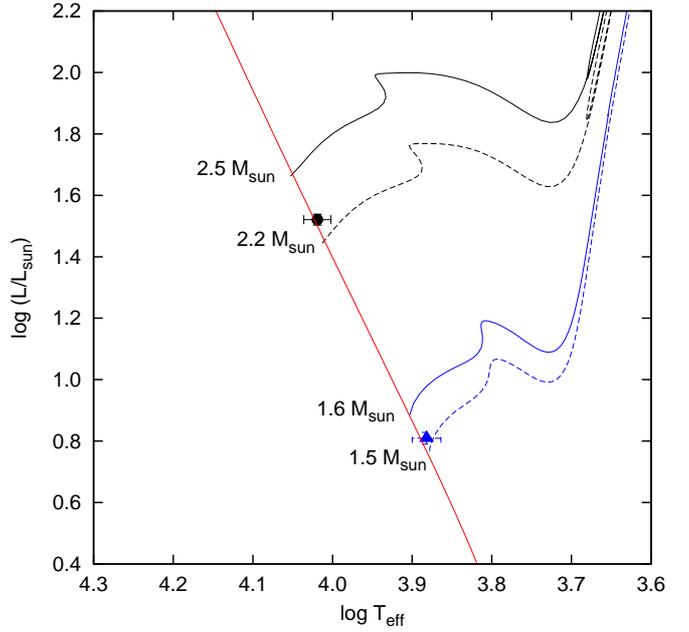}
      \caption{Positions of the primary (black filled dot) and secondary (blue filled triangle) components
in the HR diagram. The red solid line represents the ZAMS from the
\citet{gim89} models. The evolutionary tracks labelled with the corresponding
model masses are taken from \citet{ber2009}.}
         \label{evtrck}
\end{figure}

In Fig.~\ref{evtrck} we show the positions of the components in
the Hertzsprung-Russell (HR) diagram, with the ZAMS line taken from \citet{gim89} for the
solar composition. Both components are clearly located very
close to ZAMS. In the same figure, the evolutionary tracks by
\citet{ber2009} for the composition of $z$=0.017, $y$=0.3 are also shown.
These agree well with the masses of the components that we derived.

\begin{table}
    \centering
\begin{center}
\vspace{0.1cm} \caption[ ]{Basic properties of \bd.}
\label{pyhs}
\begin{tabular}{lccccccc}
 \hline\hline\noalign{\smallskip}
Element && Primary & Secondary\\
 \noalign{\smallskip}\hline\noalign{\smallskip}
$M$     &($M_{\sun}$) & $2.24\pm0.07$ & $1.52\pm0.03$\\
$R$     &($R_{\sun}$) & $1.76\pm0.01$ & $1.46 \pm0.01$\\
$T_\mathrm{eff}$ &(K) & $10450\pm420$ & $7623\pm328$\\
$log L$ &($L_{\sun}$) & $1.52\pm0.08$ & $0.81\pm0.07$\\
$M_\mathrm{bol}$        &(mag) & $0.9\pm0.2$ & $2.7\pm0.2$\\
$\lgg$  &[cgs] & $4.29\pm0.01$ & $4.29\pm0.01$\\
 $V_0$   &(mag) & $9.02\pm0.01$ & $10.44\pm0.01$\\
 (\bv)   &(mag) & $0.017\pm0.014$ & $0.316\pm0.014$\\
 BC      &(mag) & -0.310 & 0.028\\
 (\bv)$_0$      &(mag)  & $-0.037\pm0.014$ & $0.243\pm0.014$\\
 $E$(\bv)       &(mag)  & 0.054 & 0.073\\
 $M_v$  &(mag)  & $1.25\pm0.17$ & $2.69\pm0.19$\\
 $(m-M)_v$      &(mag)  & $7.77\pm0.18$ & $7.75\pm0.19$\\
 $A_v$  &(mag)  & $0.17\pm0.04$ & $0.23\pm0.04$\\
 $d$    &(pc)   & 332 & 320\\
\noalign{\smallskip}\hline\noalign{\smallskip}
\end{tabular}
\end{center}
\end{table}


 \section{Is \bd a member of the \dle?}
Although there were some doubts about the very existence of the $\delta$~Lyr (Stephenson~1) cluster,
\citet{khar2005}, \citet{khar2013} and \citet{dias2014} have included \dlc
in their open clusters (OCs) catalogues. \citet{khar2004} and
\citet{khar2013} use three different criteria for the membership
of individual stars in each considered cluster: proper motions $P_{\rm kin}$,
photometric properties $P_{\rm ph}$, and spatial properties $P_{\rm sp}$ of
the stars, respectively. For \dle, \citet{khar2013} give $E$(\bv)$=0$\m031, RV$=-27.5$~\ks,
and distance $d=373$~pc.

For an independent determination of the reddening and distance modulus of
the open cluster $\delta$~Lyr, we decided to use only
the stars that have  membership probabilities over 50~\% in each of
the three criteria used by \citet{khar2004}. They are listed in Table~\ref{mem}.
To fit the theoretical main sequence of the cluster in both colour-colour diagram (hereafter CCD) and colour-magnitude diagram (hereafter CMD),
we consider only these highly probable cluster members. The theoretical
main sequences in CCD and CMD are shifted along the axes to fit the observed main sequences.

\begin{figure}
   \centering
   \resizebox{\hsize}{!}{\includegraphics[scale=0.45,angle=270]{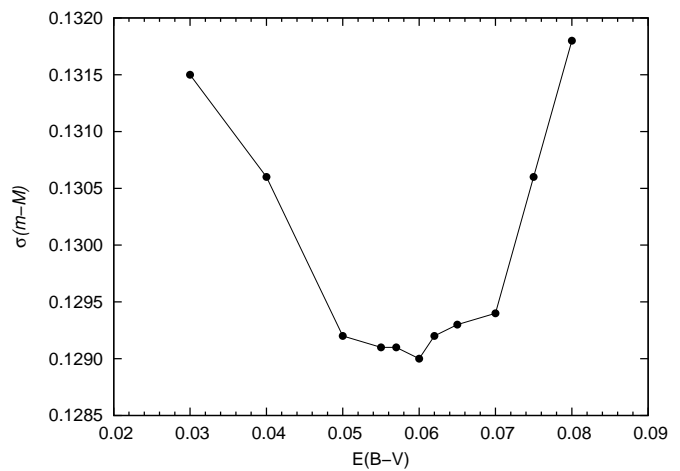}}
      \caption{Iterative determination of the optimal $E(B-V)$ -- see
 text for details.}
         \label{ebv}
\end{figure}

\begin{figure*}
   \begin{center}
   \begin{tabular}{cc}
  \hspace{-1 cm}
  {\includegraphics[scale=0.6,angle=270]{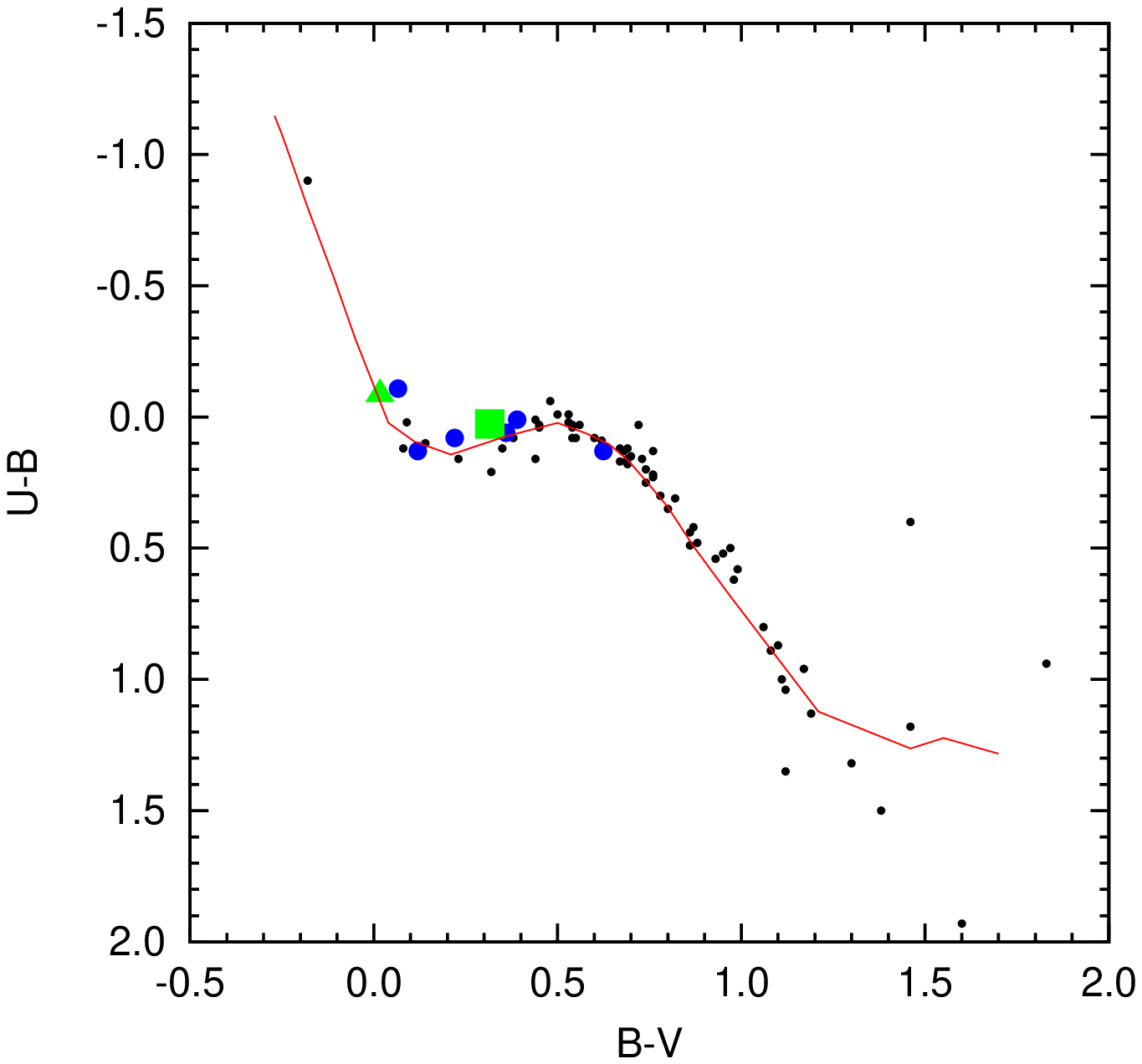}}
   \hspace{-3 cm}
   {\includegraphics[scale=0.6,angle=270]{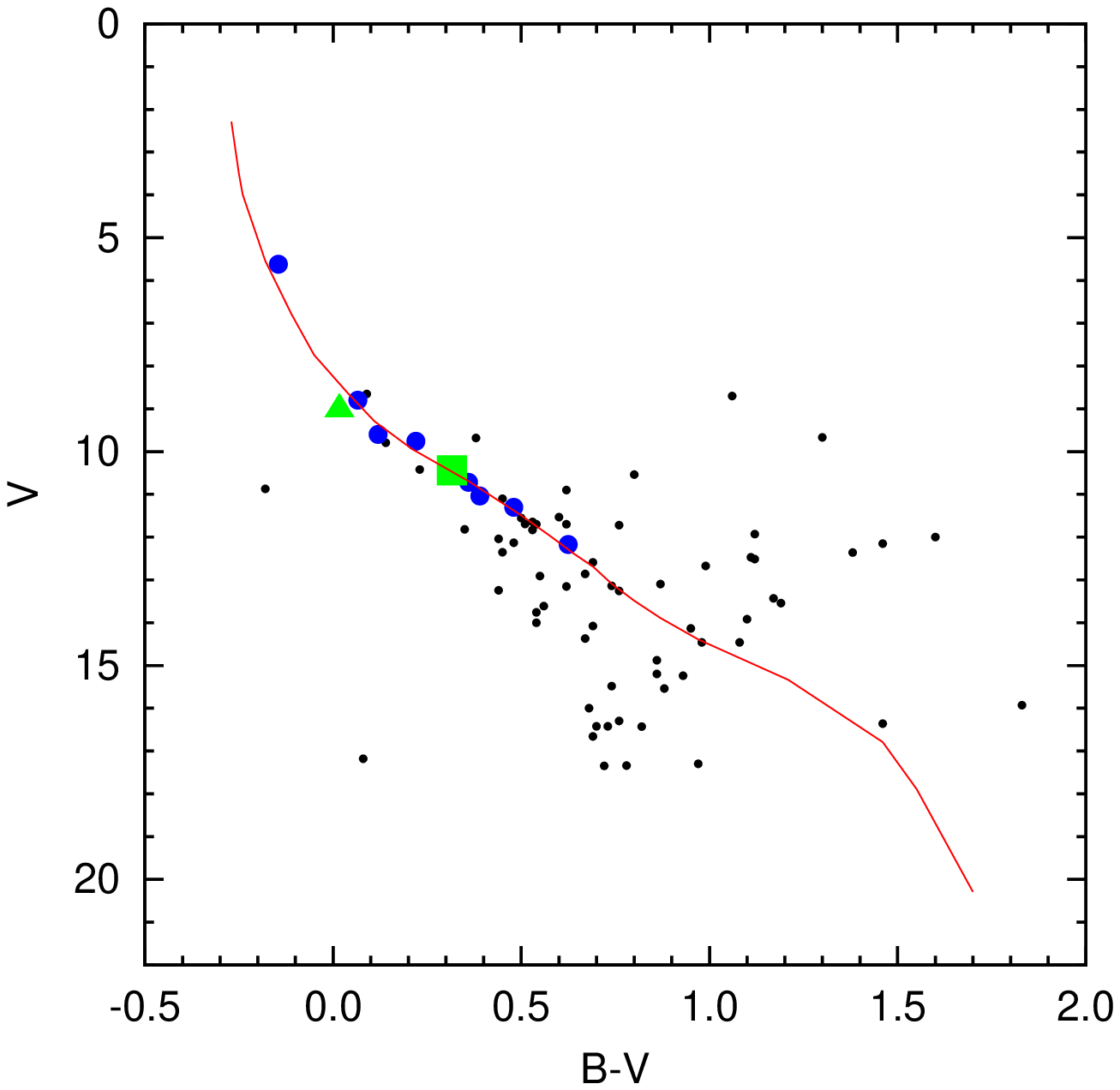}}\\
    \hspace{-2.8 cm}
   \end{tabular}
      \caption{ CCD (left panel) and CMD (right panel) diagrams for \dle.
The solid lines represent the theoretical ZAMS shifted in both axes
by appropriate amounts (see the text).The triangle and square symbols
denote the primary and secondary component, respectively. Black dots represent the stars in the field of the cluster while the big blue dots represent  the stars having high membership probability.}
   \label{HR}
   \end{center}
   \end{figure*}

We used the theoretical main sequence given by \citet{johnson66} and \citet{sch82} and represented it by a~sixth order polynomial. Then we fitted this polynomial to the most probable members of $\delta$~Lyr cluster. After this, we carried out a~searching
procedure for the colour excess $E$(\bv). To do so, we first shifted
the theoretical main sequence along the (\bv) axis for a given value of
$E$(\bv), and then determined the value of the distance modulus by hand, to find the smallest standard deviation $\sigma(m-M)$ for
the differences between the theoretical and observed main sequences.
Consequently, the resulting value of the distance modulus corresponds to
the smallest error of distance $d$ for the value of the corresponding $E$(\bv),
too. The procedure was then repeated with a new value for the reddening.
We show the variation of the smallest standard deviation $\sigma(m-M)$, which was  obtained for the cluster's distance modulus as a function of
the reddening in Fig.~\ref{ebv}. This\ shows that an~error of 0.01 mag
for $E$(\bv) is acceptable and we adopted $E$(\bv)$=0$\m$06\pm0.01$ for
the cluster. This value of the reddening corresponds to an apparent distance
modulus $(m-M)_v=7$\m$98\pm0.13$. Adopting the visual extinction
$A_v=3.1E(\bv)$, we obtain $A_v = 0.19$ mag and the distance of \dlc of $362\pm22$ pc, lower
than that found by \citet{khar2013}.
In addition, we estimate a~reddening of about 0\m04 in (\ub) from the CCD,
as shown in Fig.~\ref{HR}. In Fig.~\ref{HR}
the photometric \ubv data were taken from \citet{eggen68} and \citet{bron63}
and the stars in the vicinity of the \dlc are shown, as well as the most
probable cluster members and the primary and
secondary components of \bde.

\begin{table}
 \centering
 \caption[]{\label{mem} Most probable members of the $\delta$~Lyr cluster.}
\begin{tabular}{lccccc}
 \hline\hline\noalign{\smallskip}
WEBDA No & TYC No & $P_{kin}$ & $P_{ph}$ & $P_{sp}$ \\
 \noalign{\smallskip}\hline\noalign{\smallskip}
11 & 2650 1250 & 0.82 & 0.56 & 1.00 \\
528 & 2650 1237 & 0.69 & 1.00 & 1.00 \\
529 & 2650 2146 & 0.93 & 1.00 & 1.00 \\
533 & 2651 1056 & 0.98 & 0.81 & 1.00\\
542\tablefootmark{*} & 2651 802 & 0.67 & 0.91 & 1.00 \\
545 & 2651 882 & 0.94 & 0.73 & 1.00 \\
96 & 2651 772 & 0.76 & 1.00 & 1.00 \\
552 & 2651 189 & 0.99 & 1.00 & 1.00 \\
\hline\noalign{\smallskip}
\tablefoottext{*}{\bde.}
\end{tabular}
\end{table}

\begin{table}
\addtolength{\tabcolsep}{-2.5pt}
 \caption[]{\label{memrv} RV measurements of stars in \dle.}
\begin{tabular}{lccc}
 \hline\hline\noalign{\smallskip}
  Name &
  RV &
  References &
  Notes \\
 \noalign{\smallskip}\hline\noalign{\smallskip}
$HD 174959$&$-12.4\pm2.4$&1&--\\
$HD 175081$&$-26.0\pm7.4$&2&--\\
$\delta_1$ Lyr&$-17.2\pm4.3$&1&spectroscopic binary\\
$\delta_2$ Lyr&$-25.5\pm0.5$&1&pulsating star\\
\bd&$-17.8\pm0.5$&this study&EB\\
 \noalign{\smallskip}\hline\noalign{\smallskip}
\end{tabular}
\tablebib{(1) \citet{gont2006}; (2) \citet{khar2007}.
}
\end{table}

\citet{khar2005} give the cluster radius as 0.87 degrees and
RV$= -21.6$~\kms \citep[compared with -27.5 \kms in][]{khar2013}. Only four stars have the RV measurements in the \dlc field. These stars and the RV measurements are given in Table~\ref{memrv}.
Low membership probabilities are given for $\delta_2$~Lyr and HD~174959
in \citep{khar2013}. However the probabilities of the membership
for $\delta_1$~Lyr and HD~175081 are higher. The systemic RV of \bd from our solution agrees with $\delta_1$~Lyr, but not
with HD~175081. Nevertheless, the locations of the components of \bd in
both CCD and CMD (see Fig.~\ref{HR}), are in good agreement with the other
highly probable members of the cluster. Also, \citet{khar2004,khar2013}
classify the system as a highly probable member. The adopted distance,
which is the luminosity-weighted average of the distances obtained for the components
from Table~\ref{pyhs}, is $330\pm29$ pc, consistent with the cluster's distance ($362\pm22$ pc)
within the quoted errors. So, we conclude that \bd is one of the most
probable members of the \dle, although obtaining more
observations, in particular RVs, for more cluster member candidates is
still  very desirable.

\begin{figure}
   \centering
    \hspace*{-1.5cm}
    {\includegraphics[scale=0.65,angle=270]{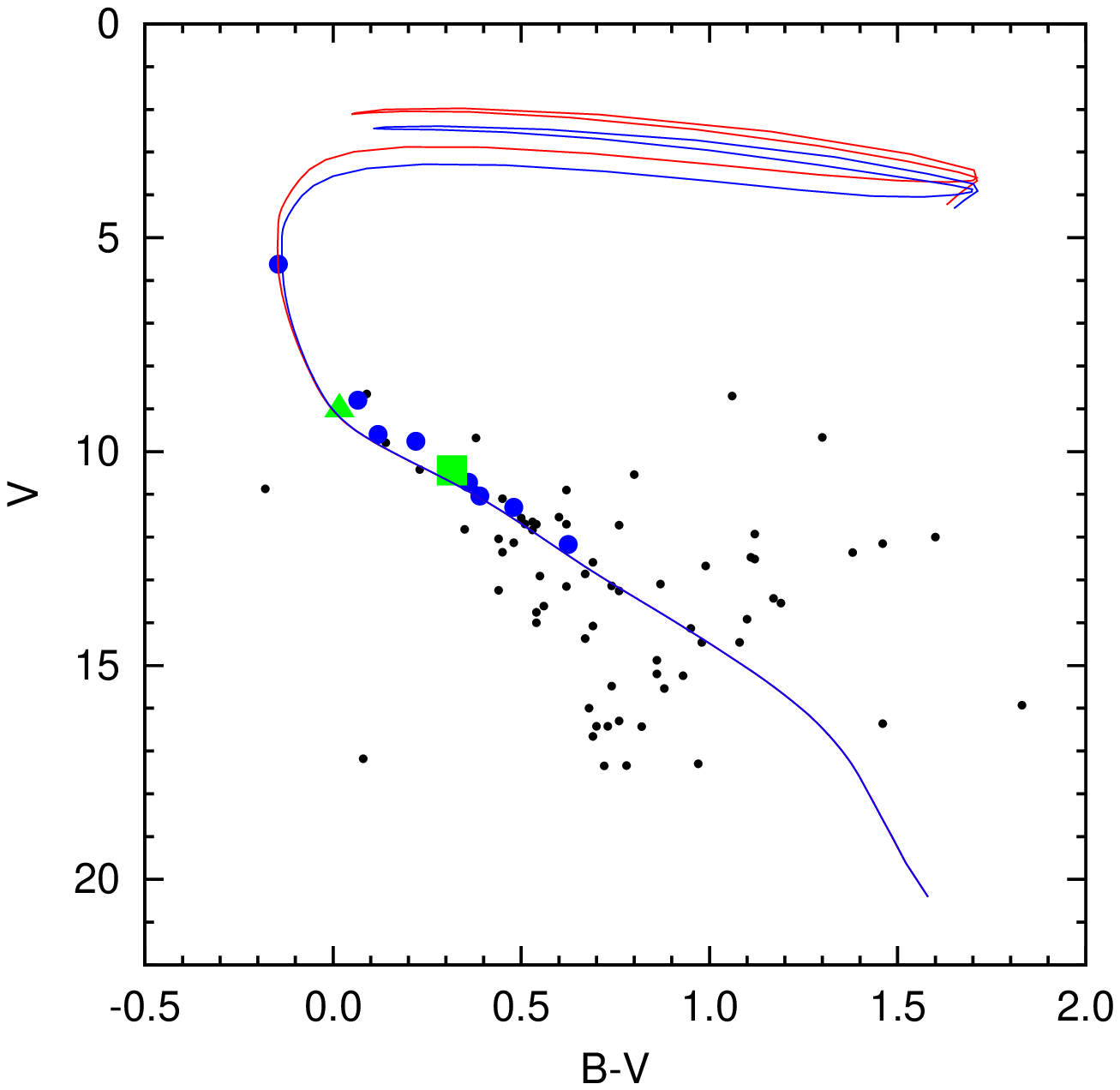}}
      \caption{ HR diagram of \dle. The  triangles and squares
denote the primary and secondary components, respectively. Black dots represent the stars in the field of the cluster and the big blue dots, the stars having high membership probability. Padova isochrones taken from
\citet{ber2009} are shown for the reddening of 0\m06 and the apparent
distance modulus of 7\m98. The isochrones shown are for 25 (red line)
and 32 (blue line)~Myrs and for the chemical composition
$(y,z) = (0.3,0.017)$.}
         \label{iso}
\end{figure}

At  this point we decided to estimate the age of the cluster. So we tried to obtain best isochrone fitting for the cluster by using only the most probable members given in Table~\ref{mem}.
To do so, we considered both the turn-off point and the slope of
the main sequence together. We compared our CMD with the YZVAR Padova
Isochrones produced by \citet{ber2009}. We obtained the best fit
for the isochrone of $(y,z) = (0.3, 0.017)$ for the cluster.
Finally, as seen in Fig.~\ref{iso}, the age of the cluster is
between $\log t = 7.4$ and 7.5. So we accepted an~age of about
$3\times10^7$ yrs for the cluster. This value is also in  good accordance with the evolutionary status of \bde, as  pointed out in Sec. 3.7.

 \section{Conclusion}
Obtaining and studying the first series of electronic spectra of \bd and
solving the RV curves of both binary components together with already
published light curves, we obtained the first complete set
of basic physical properties for this eclipsing binary,
which is a probable member of \dle. The results can be summarised as follows.

\begin{enumerate}
\item The masses, radii, effective temperatures, and absolute luminosities
summarised in Table~\ref{pyhs} show that both components of \bd are
little evolved from the ZAMS and compatible with the published
evolutionary models of \citet{ber2009}.
\item The spin-orbit synchronisation for our solution would predict
the projected rotational velocities of 20.7 and 17.2~\kms for the
primary and secondary, respectively. This does not
contradict the values we estimated from the line profiles, taking
their associated errors into consideration.
\item We reinforce the conclusion that \bd is a probable member of the
\dle. Both binary components seem to agree with the other members of \dlc on the main sequence in CMD.
\item Using the de-reddened magnitudes from our solution individually for
both components, we derived the weighted mean distance of \bd of
$330\pm29$~pc. This distance is within the
boundaries of the cluster distance estimated from CMD of \dlc and, if the
membership of \bd in the cluster can be confirmed definitively, it would
represent the most accurate estimate for the distance of the \dle.
\end{enumerate}

\begin{acknowledgements}
We thank to Dr. O. \"{O}zdarcan for sharing his photometric data of \bd
with us, and to Dr.~\"O.~\c Cak{\i}rl{\i} for his support concerning the program {\tt COMPO2} for temperature searching process with the composite spectra method.
The visit of EK in Praha and Ond\v{r}ejov was supported by The Scientific and Technological Research Council of Turkey (T\"{U}B{\.I}TAK) by the fellowship of B{\.I}DEP-2214 International
Doctoral Research Fellowship Programme.
She thanks colleagues in the Astronomical Institute of the Charles
University and Astronomical Institute of the Academy of Sciences of
the Czech Republic for their hospitality and help during her stay.
Research of PH, MW, and JN was
supported by grants P209/10/0715 and GA15-02112S of the Czech Science
Foundation and the research of JN was additionally supported with
grant no.~250015 of Grant Agency of the Charles University in Prague.
We benefited from the use of the SIMBAD database and the VizieR service
operated at CDS, Strasbourg, France and the NASA's Astrophysics Data System
Bibliographic Services and the WEBDA open cluster database.
\end{acknowledgements}

\bibliographystyle{aa}
\bibliography{evrim}

\begin{thebibliography}{39}
\expandafter\ifx\csname natexlab\endcsname\relax\def\natexlab#1{#1}\fi

\bibitem[{{Anthony-Twarog}(1984)}]{ant84}
{Anthony-Twarog}, B.~J. 1984, \aj, 89, 655

\bibitem[{{Bertelli} {et~al.}(2009){Bertelli}, {Nasi}, {Girardi}, \&
  {Marigo}}]{ber2009}
{Bertelli}, G., {Nasi}, E., {Girardi}, L., \& {Marigo}, P. 2009, \aap, 508, 355

\bibitem[{{Bronkalla}(1963)}]{bron63}
{Bronkalla}, W. 1963, Astronomische Nachrichten, 287, 249

\bibitem[{{Castelli} \& {Kurucz}(2004)}]{cas2004}
{Castelli}, F. \& {Kurucz}, R.~L. 2004, ArXiv Astrophysics e-prints, 0405087

\bibitem[{{Claret}(1998)}]{clar98}
{Claret}, A. 1998, \aaps, 131, 395

\bibitem[{{Claret}(2001)}]{clar2001}
{Claret}, A. 2001, \mnras, 327, 989

\bibitem[{{Claret} \& {Gim{\'e}nez}(1989)}]{gim89}
{Claret}, A. \& {Gim{\'e}nez}, A. 1989, \aaps, 81, 1

\bibitem[{{de Laverny} {et~al.}(2012){de Laverny}, {Recio-Blanco}, {Worley}, \&
  {Plez}}]{delaverny2012}
{de Laverny}, P., {Recio-Blanco}, A., {Worley}, C.~C., \& {Plez}, B. 2012,
  \aap, 544, A126

\bibitem[{{Dias} {et~al.}(2014){Dias}, {Monteiro}, {Caetano}, {L{\'e}pine},
  {Assafin}, \& {Oliveira}}]{dias2014}
{Dias}, W.~S., {Monteiro}, H., {Caetano}, T.~C., {et~al.} 2014, \aap, 564, A79

\bibitem[{{Eggen}(1968)}]{eggen68}
{Eggen}, O.~J. 1968, \apj, 152, 77

\bibitem[{{Eggen}(1972)}]{eggen72}
{Eggen}, O.~J. 1972, \apj, 173, 63

\bibitem[{{Eggen}(1983)}]{eggen83}
{Eggen}, O.~J. 1983, \mnras, 204, 391

\bibitem[{{Frasca} {et~al.}(2006){Frasca}, {Guillout}, {Marilli}, {Freire
  Ferrero}, {Biazzo}, \& {Klutsch}}]{Fras2006}
{Frasca}, A., {Guillout}, P., {Marilli}, E., {et~al.} 2006, \aap, 454, 301

\bibitem[{{Gontcharov}(2006)}]{gont2006}
{Gontcharov}, G.~A. 2006, Astronomy Letters, 32, 759

\bibitem[{{Gray}(2005)}]{gra2005}
{Gray}, D.~F. 2005, The Observation and Analysis of Stellar Photospheres, 3rd
  Edition, UK: Cambridge University Press, 533 pp

\bibitem[{{Hadrava}(1995)}]{had95}
{Hadrava}, P. 1995, \aaps, 114, 393

\bibitem[{{Hadrava}(1997)}]{had97}
{Hadrava}, P. 1997, \aaps, 122, 581

\bibitem[{{Hadrava}(2004)}]{had2004}
{Hadrava}, P. 2004, Publications of the Astronomical Institute of the
  Czechoslovak Academy of Sciences, 92, 15

\bibitem[{{Hadrava}(2009)}]{had2009}
{Hadrava}, P. 2009, ArXiv e-prints, \# 0909.0172

\bibitem[{{Harmanec} {et~al.}(2015){Harmanec}, {Koubsk{\'y}}, {Nemravov{\'a}},
  {Royer}, {Briot}, {North}, {Lampens}, {Fr{\'e}mat}, {Yang}, {Bo{\v z}i{\'c}},
  {Kotkov{\'a}}, {{\v S}koda}, {{\v S}lechta}, {Kor{\v c}{\'a}kov{\'a}},
  {Wolf}, \& {Zasche}}]{zarfin30}
{Harmanec}, P., {Koubsk{\'y}}, P., {Nemravov{\'a}}, J.~A., {et~al.} 2015, \aap,
  573, A107

\bibitem[{{Horn} {et~al.}(1996){Horn}, {Kub\'at}, {Harmanec}, {Koubsk\'y},
  {Hadrava}, {\v{S}imon}, {\v{S}tefl}, \& {\v{S}koda}}]{sef0}
{Horn}, J., {Kub\'at}, J., {Harmanec}, P., {et~al.} 1996, \aap, 309, 521

\bibitem[{{Johnson}(1966)}]{johnson66}
{Johnson}, H.~L. 1966, \araa, 4, 193

\bibitem[{{Kharchenko} {et~al.}(2004){Kharchenko}, {Piskunov}, {R{\"o}ser},
  {Schilbach}, \& {Scholz}}]{khar2004}
{Kharchenko}, N.~V., {Piskunov}, A.~E., {R{\"o}ser}, S., {Schilbach}, E., \&
  {Scholz}, R.-D. 2004, Astronomische Nachrichten, 325, 740

\bibitem[{{Kharchenko} {et~al.}(2005){Kharchenko}, {Piskunov}, {R{\"o}ser},
  {Schilbach}, \& {Scholz}}]{khar2005}
{Kharchenko}, N.~V., {Piskunov}, A.~E., {R{\"o}ser}, S., {Schilbach}, E., \&
  {Scholz}, R.-D. 2005, \aap, 438, 1163

\bibitem[{{Kharchenko} {et~al.}(2013){Kharchenko}, {Piskunov}, {Schilbach},
  {R{\"o}ser}, \& {Scholz}}]{khar2013}
{Kharchenko}, N.~V., {Piskunov}, A.~E., {Schilbach}, E., {R{\"o}ser}, S., \&
  {Scholz}, R.-D. 2013, \aap, 558, A53

\bibitem[{{Kharchenko} {et~al.}(2007){Kharchenko}, {Scholz}, {Piskunov},
  {R{\"o}ser}, \& {Schilbach}}]{khar2007}
{Kharchenko}, N.~V., {Scholz}, R.-D., {Piskunov}, A.~E., {R{\"o}ser}, S., \&
  {Schilbach}, E. 2007, Astronomische Nachrichten, 328, 889

\bibitem[{{Nasseri} {et~al.}(2014){Nasseri}, {Chini}, {Harmanec}, {Mayer},
  {Nemravov{\'a}}, {Dembsky}, {Lehmann}, {Sana}, \& {Le Bouquin}}]{nasseri2014}
{Nasseri}, A., {Chini}, R., {Harmanec}, P., {et~al.} 2014, \aap, 568, A94

\bibitem[{{{\"O}zdarcan} {et~al.}(2012){{\"O}zdarcan}, {Sipahi}, \&
  {Dal}}]{ozdar2012}
{{\"O}zdarcan}, O., {Sipahi}, E., \& {Dal}, H.~A. 2012, \na, 17, 483

\bibitem[{{Palacios} {et~al.}(2010){Palacios}, {Gebran}, {Josselin}, {Martins},
  {Plez}, {Belmas}, \& {L{\`e}bre}}]{palacios2010}
{Palacios}, A., {Gebran}, M., {Josselin}, E., {et~al.} 2010, \aap, 516, A13

\bibitem[{{Pr{\v s}a} \& {Zwitter}(2005)}]{prsa2005}
{Pr{\v s}a}, A. \& {Zwitter}, T. 2005, \apj, 628, 426

\bibitem[{{Pr{\v s}a} \& {Zwitter}(2006)}]{prsa2006}
{Pr{\v s}a}, A. \& {Zwitter}, T. 2006, \apss, 36

\bibitem[{{Rossiter}(1924)}]{rossit}
{Rossiter}, R.~A. 1924, \apj, 60, 15

\bibitem[{{Schmidt-Kaler}(1982)}]{sch82}
{Schmidt-Kaler}, T., ed. 1982, {Landolt-B{\"o}rnstein: Numerical Data and
  Functional Relationships in Science and Technology - New Series ''
  Gruppe/Group 6 Astronomy and Astrophysics '' Volume 2 Schaifers/Voigt:
  Astronomy and Astrophysics / Astronomie und Astrophysik '' Stars and Star
  Clusters / Sterne und Sternhaufen}

\bibitem[{{\v{S}koda}(1996)}]{sko96}
{\v{S}koda}, P. 1996, in Astronomical Society of the Pacific Conference Series,
  Vol. 101, Astronomical Data Analysis Software and Systems V, ed. G.~H.
  {Jacoby} \& J.~{Barnes}, 187

\bibitem[{{Stephenson}(1959)}]{Step59}
{Stephenson}, C.~B. 1959, \pasp, 71, 145

\bibitem[{{{\v S}koda} \& {Hadrava}(2010)}]{sko2010}
{{\v S}koda}, P. \& {Hadrava}, P. 2010, in Astronomical Society of the Pacific
  Conference Series, Vol. 435, Binaries - Key to Comprehension of the Universe,
  ed. A.~{Pr{\v s}a} \& M.~{Zejda}, 71

\bibitem[{{Valdes} {et~al.}(2004){Valdes}, {Gupta}, {Rose}, {Singh}, \&
  {Bell}}]{val2004}
{Valdes}, F., {Gupta}, R., {Rose}, J.~A., {Singh}, H.~P., \& {Bell}, D.~J.
  2004, \apjs, 152, 251

\bibitem[{{Violat-Bordonau}(2008)}]{vio2008}
{Violat-Bordonau}, T., F.~.-H. 2008, Information Bulletin on Variable Stars,
  5900, 7

\bibitem[{{Wilson} \& {Devinney}(1971)}]{wd71}
{Wilson}, R.~E. \& {Devinney}, E.~J. 1971, \apj, 166, 605

\end{thebibliography}

\end{document}